# Evidence of spin-density-wave order in RFeAsO from measurements of thermoelectric power

M. Matusiak[1], T. Plackowski[1], Z. Bukowski[2], N. D. Zhigadlo[2], and J. Karpinski[2]

1. Institute of Low Temperature and Structure Research, Polish Academy of Sciences,
P.O. Box 1410, 50-950 Wrocław, Poland
2. Laboratory for Solid State Physics, ETH Zurich, 8093 Zurich, Switzerland

**Abstract**

Data on the magneto-thermopower and specific heat of three compounds belonging to "1111" oxypnictides family are reported. One specimen (SmAsFeO$_{0.8}$F$_{0.2}$) is a superconductor with $T_c$ = 53 K, while two others (SmAsFeO and NdAsFeO) are nonsuperconducting parent compounds. Our results confirm that spin density wave (SDW) order is present in SmAsFeO and NdAsFeO. In these two samples a strict connection between the thermoelectric power and electronic specific heat is found in the vicinity of SDW transition, what indicates that the chemical potential of charge carriers strongly depends on temperature in this region. Low temperature data suggest presence of significant contribution magnon-drag to the thermoelectric power.

**Text**

Intense investigations of properties of the rare earth iron-oxypnictides reveal some similarities between their phase diagram and that of the cuprate superconductors. Namely, in the cuprates superconductivity emerges when mobile 'electrons' or 'holes' are doped into antiferromagnetic parent compounds[1,2], and an analogous behavior is observed in the iron-based superconductors, where an electron doping seems to suppress the spin density wave (SDW) instability allowing superconductivity to appear[3,4]. However, there were raised some doubts, based on results of neutron diffraction studies, whether this behavior is common for all iron oxypnictides[5]. Clarifying this uncertainty can be important in determining a role that is played by magnetic interactions in the mechanism of superconductivity. Since there is no obvious way to distinguish between itinerant and localized magnetism[6], we utilize an experimental technique that allows the study of magnetic ordering of charge carriers. In order to do it, we study the specific heat and thermoelectric power (measured in the magnetic field up to 13 T) of doped and undoped oxypnictides, including a NdFeAsO compound that caused above mentioned doubts.

Polycrystalline samples of NdFeAsO and SmFeAsO were prepared by conventional solid state reaction. First, NdAs, SmAs and FeAs were synthesized from pure elements in evacuated silica ampoules at 600 ºC. In the next step stoichiometric amounts of NdAs, (or SmAs), and FeO were weighed and thoroughly mixed. The raw mixtures were pressed into pellets, wrapped in Ta foil and sealed in silica tubes under reduced pressure of Ar gas. The pellets were heated at 1160 ºC for 40 hours with intermittent regrinding and pelletizing. A high-density sample of $SmFeAsO_{0.8}F_{0.2}$ were prepared by a high-pressure and high-temperature method using a cubic anvil apparatus[7]. The stoichiometric mixture of SmAs, FeAs, Fe, $Fe_2O_3$, and $SmF_3$ was placed in a BN container inside a pyrophyllite cube equipped with a graphite heater. The compound was synthesized at a pressure of 3 GPa and temperature 1350 ºC for 4.5 h. The phase purity of the obtained samples was checked by means of powder X-ray diffraction (XRD) carried out on a STOE diffractometer using Cu-Kα radiation and a graphite monochromator. The XRD patterns of NdFeAsO and SmFeAsO samples showed no detectable amount of impurities while $SmFeAsO_{0.8}F_{0.2}$ contained some amount of Sm oxyfluoride. The lattice parameters calculated from XRD data were a=3.965 Å and c=8.575 Å for NdFeAsO, a=3.937 Å and c=8.500 Å for SmFeAsO, a=3.927 Å and c=8.461 Å for $SmFeAsO_{0.8}F_{0.2}$.

The magnetic field in the magneto-thermopower measurement was parallel to the temperature gradient. A sample was clamped between two spring-loaded copper blocks provided with heaters and a pair of thermometers (Cernox 1050). The blocks were thermally insulated from the surrounding, therefore a thermal difference of any sign might be produced by the heaters. The voltage difference between blocks was measured using an A20 (EM Electronics) low-noise preamplifier. More details about the method can be found in Ref. [8].

The specific heat ($C_p$) was measured using a heat-flow calorimeter[9]. In this method the sample is connected with a heat sink by means of a sensitive heat-flow meter of high thermal conductance. To sense the heat flux we used a commercial, miniature, one-stage Peltier cell with sensitivity of 0.45 V/W at room temperature and 0.08 V/W at liquid nitrogen temperature. The sample was fixed on the cell top plate, made of 0.5-mm-thick alumina. The bottom of the heat-flow meter was permanently attached to the heat sink (a massive copper block) of controlled temperature. An in-field calibrated Pt thermometer was attached to the sink. Such a device was surrounded by a double passive radiation screen (gold plated). Both screens were in a good thermal contact with the sink. The whole ensemble was evacuated down to $10^{-6}$ hPa and placed in the gas-flow variable-temperature insert of an Oxford cryostat with a 13/15 T superconducting magnet.

First we present data on a temperature dependence of the specific heat ($C_p$) of the superconducting $SmAsFeO_{0.8}F_{0.2}$ sample (Fig. 1 (*a*)) accompanied by $C_p(T)$ dependences taken on specimens of two non-superconducting parent compounds: $SmAsFeO$ (Fig. 1 (*b*)) and $NdAsFeO$ (Fig. 1 (*c*)). In the fluorine doped $SmAsFeO_{0.8}F_{0.2}$ sample, we see a kink at the $T = 53$ K that is related to the formation of the superconducting state. An application of the magnetic field of 13 T almost completely smears the kink out, while a hump near $T_{SDW} \approx 140$ K in both parent compounds is resistant to an influence of the magnetic field. This anomaly can be associated with the SDW transition[10,11], but on the other hand there is another, structural, transition (from tetragonal to orthorhombic system) that occurs at the same or very near temperature[12,13,14]. In Fig. 2 there are presented data on the thermoelectric power (TEP) versus temperature for: $SmAsFeO_{0.8}F_{0.2}$ (Fig. 2 (*a*)), $SmAsFeO$ (Fig. 2 (*b*)) and $NdAsFeO$ (Fig. 2 (*c*)) samples. A high temperature part of the $S(T)$ dependences look similar for all three samples: $S$ is negative, has a positive slope and changes almost linearly with $T$, while at low temperatures the thermopower of parent compounds ($SmAsFeO$ and $NdAsFeO$) develops a broad maximum below $T \sim 200$ K. A qualitatively the same type of the thermopower behavior has been reported for $LaFeAsO$[15], and we are convinced that the maximum appears due to formation of the SDW state, since such a behavior has been already described theoretically[16] and observed in many other compounds[17,18,19]. However, we think that more interesting result of influence of the emerging SDW order on the thermopower reveals, when the TEP data is differentiated and presented as $TdS/dT$. In Fig. 3 the thermoelectric power is shown in this way, along with a temperature dependence of electronic specific heat ($C_e$). We assume that in the vicinity of the transition $C_e$ consists of an excess specific heat and a background that varies slowly with temperature. Therefore, for both samples $C_e$ was estimated by creating a curve of the Einstein and Debye components fitted to the measured data well outside the SDW transition region and subtracting the obtained fit from $C_p(T)$ curve. Because size of the missed background does not affect our analysis, it is assumed to be zero for the sake of simplicity. Connection between $C_e(T)$ and $TdS/dT$ dependencies in the vicinity of itinerant magnetic transitions has been already noted for instance in nickel[20] or chromium[21], however its origin was judged by some researchers as debatable[22,23]. Our considerations begin with realization that the thermopower should be properly defined in terms of the electrochemical ($\bar{\mu}$), rather than only electrical ($\psi$), potential[24]:

$$S\left[\frac{\mu V}{K}\right] \equiv -\frac{\nabla \bar{\mu}}{ne\nabla T}, \qquad (1)$$

Where: $e$ is charge of the electron, and $n$ is charge carriers concentration that incorporates their sign. Because $\bar{\mu} \equiv \mu + e\psi$ ($\mu$ denotes chemical potential), we can write:

$$-S = \frac{1}{ne}\frac{d\mu}{dT} + \frac{1}{n}\frac{d\psi}{dT}. \qquad (2)$$

There can be seen now two contributions to the thermoelectric power: the usual one related to the electrical potential difference caused by thermal diffusion of charge carriers ($S_{diff} = -\frac{1}{n}\frac{d\psi}{dT}$), and one arising from temperature dependence of the chemical potential ($S_\mu = -\frac{1}{ne}\frac{d\mu}{dT}$). The latter component usually can be neglected, but it could manifests itself in rare cases, when the chemical potential strongly depends on temperature. The Gibbs-Duhem equation for the charge carriers in metal states that:

$$d\mu = -\hat{S}dT + vdp - MdH, \qquad (3)$$

where $\hat{S}$ is the electron gas entropy (the caret is added to distinguish it from $S$ that denotes TEP), $v$ and $p$ are volume and pressure of the electron gas, M is magnetization, and H is magnetic field. If the second and third terms could be omitted (because the pressure of the electron gas is modified only through the lattice volume, which changes are small, and an experiment is carried at constant magnetic field), we have:

$$\frac{d\mu}{dT} = -\hat{S} = -\left(\int_0^T \frac{C_e}{T}dT\right). \qquad (4)$$

It means that $S_\mu$ is a direct measure of the electron entropy:

$$S_\mu = -\frac{1}{ne}\hat{S}. \tag{5}$$

Thus the equations 4 and 5 can be rewritten as:

$$ne\frac{dS_\mu}{dT} = \frac{C_e}{T}. \tag{6}$$

This is formula of great importance, since it shows that in an electronic system, which undergoes a transition causing changes of the chemical potential, the first derivative of $S_\mu$ multiplied by $T$ is simply proportional to the electronic specific heat. If the diffusion contribution to the thermopower weakly depends on temperature, we have:

$$T\frac{dS}{dT} \approx \frac{1}{ne}C_e, \tag{7}$$

what means that measurement of the thermopower is a convenient tool to study an itinerant-electron magnetism for example, where a shift of the chemical potential is expected to occur at the transition temperature. In the figure 4 there are presented parametric plots of $C_e$ versus $-TdS/dT$ for SmAsFeO and NdAsFeO. We find both quantities related proportionally on both sides of the transition and the slope $\frac{1}{ne}$ was used to determine the concentration of charge carriers. Values of $n$ turn out to be similar above and below $T_{SDW}$ ($n_{fu} \approx -0.38$ carrier per formula unit for SmAsFeO, and $n_{fu} \approx -0.26$ carrier per formula unit for NdAsFeO), what may be surprising since the SDW state should presumably gaps most of the Fermi surface. On the other hand our results support a theory introduced recently by I. I. Mazin and M.D. Johannes[25] that emphasizes a role of spin dynamics in physics of the oxypnictides. Authors conclude that on the onset of the SDW we should expect a rapid change of the electron and hole relaxation time ratio, rather than sharp drop in the carrier concentration.

We note also that $S(T)$ dependences in the parent compounds start to deviate from their high-temperature linearity in temperatures significantly higher (~80 K) than $T_{SDW}$ and we think that it might suggest a presence of SDW fluctuations. A wide temperature range of their existence can be related to layered structure of oxypnictides that lower an effective dimensionality of the system. An analogous behavior can been seen in cuprate superconductors where a recent work on Nd-LSCO crystals[26] identifies the pseudogap phase (present below $T^*$) as a phase with fluctuating "stripe" order that is fully established at $T \sim 0.5\ T^*$. One more example is ZrTe$_3$, which has quasi 1D and 3D Fermi surfaces. A recent results of angle-resolved photoemission spectroscopy studies reported by Yokoya et al.[27], suggest that a Charge Density Wave gap in the electronic density of states opens at temperatures much higher than $T_{CDW}$ due to CDW fluctuations.

A low temperature TEP of all three samples is susceptible to the magnetic field (see insets in Fig. 2), though an analysis of normal-state behavior in the SmAsFeO$_{0.8}$F$_{0.2}$ is difficult due to the superconducting transition at $T_c$ = 53 K. The field dependent parts of the thermopower for both parent compounds are shown as [$S(B,T)$ - $S(0,T)$] plots in insets in Fig. 5. In principle the determined dependences might be a result of modification of the SDW state by an applied magnetic field[28,29], but we do not see much influence of $B$ neither on $C_p$ nor $S$ in the vicinity of $T_{SDW}$. Therefore we conclude that the magnetic field of 13 T does not alter a structure of SDW substantially. Changes of value of TEP at the low-temperature minimum ($T$ ~ 50K) in function of $B$ are presented in insets in Fig. 2, panel (*b*) and (*c*). The linear dependence of the thermopower on the magnetic field observed at low temperatures might suggest that we see a contribution from the phonon-drag thermopower ($S_{ph-drag}$), since the $S_{ph-drag}$ term is usually sensitive to the magnetic field[30] and a theory by Gurevich et al.[31] predicts that $S_{ph-drag}$ should grow linearly with *B*. Although, it would be difficult to explain a significant size of the effect within this scenario, because we study the polycrystalline samples, where it is unlikely to see a large contribution from the phonon-drag thermopower due to high amount of crystalline defects. Therefore we can suspect that another contribution to TEP, namely the magnon-drag thermopower ($S_{mag-drag}$), is considerable. It is known to behave in many aspects in similar way to phonon-drag one[32,33], but a difference between them is that the magnon-drag thermopower is much less sensitive for non-magnetic defects[34,35]. We conclude then that a field dependent contribution to the thermoelectric power could come from the magnon-drag term. Because one can anticipate spin excitations to be present in SDW state and their existence in the parent phase of pnictide superconductors has been already reported[36], we think that a magnon-drag scenario is probable. However, a definitive distinction between phonon- and magnon-drag thermopower is not trivial, since phonons and antiferromagnetic magnons at low *T* follow the same $T^3$ temperature dispersion[37]. In the Fig. 5 we present in the log-log scale a difference between *S* measured in presence and absence of the magnetic field. Plotted points represent $c*S_{mag-drag}(T)$ dependences (*c* is a constant proportional to *B*) that are drawn under assumption that dependence of diffusion thermopower on *B* is negligible. The low-temperature part of the plot of both samples follows the power $T^\alpha$ law, but $\alpha$ is equal to $2/3$, rather than 3 (what is expected for electron scattering on phonons or antiferromagnetic magnons) or $3/2$ (ferromagnetic magnons). This puzzle may need further investigation.

In conclusion, the thermoelectric power and specific heat of three rare earth iron oxypnictides (SmAsFeO$_{0.8}$F$_{0.2}$, SmAsFeO, SmAsFeO) in the magnetic field up to 13 T have been studied. The electronic specific heat in the vicinity of $T_{SDW}$ is strictly correlated with the first derivative of the thermopower $\left(T\dfrac{dS}{dT}\right)$ indicating an itinerant nature of the transition and dramatic changes that happens in the electronic structure. It is clear evidence of Spin Density Wave order in the nonsuperconducting, undoped, parent samples. A possibility of significant contribution to the thermopower from magnon-drag effect at low temperature is discussed.

**Figure Captions**

**1.** Temperature dependences of the specific heat divided by $T$ for SmAsFeO$_{0.8}$F$_{0.2}$ (*a*), SmAsFeO (*b*) and NdAsFeO (*c*). Insets show data taken in the applied magnetic field of 13 T (upper line) in a temperature region of the superconducting (panel (*a*)) or SDW (panels (*b*) and (*c*)) transition. Data taken in the magnetic field are shifted vertically for clarity.

**2.** Temperature dependences of the thermoelectric power of SmAsFeO$_{0.8}$F$_{0.2}$ (*a*), SmAsFeO (*b*) and NdAsFeO (*c*) in the applied magnetic field of 13, 5 and 0 T. Inset in panel (*a*) shows an enlarged region in the vicinity of the superconducting transition, while insets in panel (*b*) and (*c*) present field dependences of the thermopower at the low-temperature minimum.

**3.** Temperature dependences of the electronic specific heat (left axis) and derivative of thermoelectric power multiplied by -T $\left(\text{i.e. } -T\dfrac{dS}{dT}\right)$ (right axis) of SmAsFeO (*a*) and NdAsFeO (*b*).

**4.** Parametric plots of the electronic specific heat versus $-T\dfrac{dS}{dT}$ for SmAsFeO (*a*) and NdAsFeO (*b*). Solid points present data for $T > T_{\text{SDW}}$, whereas open points are for $T < T_{\text{SDW}}$. Solid lines represent linear fits.

**5.** A temperature dependence of the differences between *S* measured in presence and of absence the magnetic field of 13 T (presented as absolute values in the log-log scale). Insets show the same quantity (along with an analogous plot for $B = 5$ T) in linear scale for SmAsFeO (*a*) and NdAsFeO (*b*).

**Fig. 1**

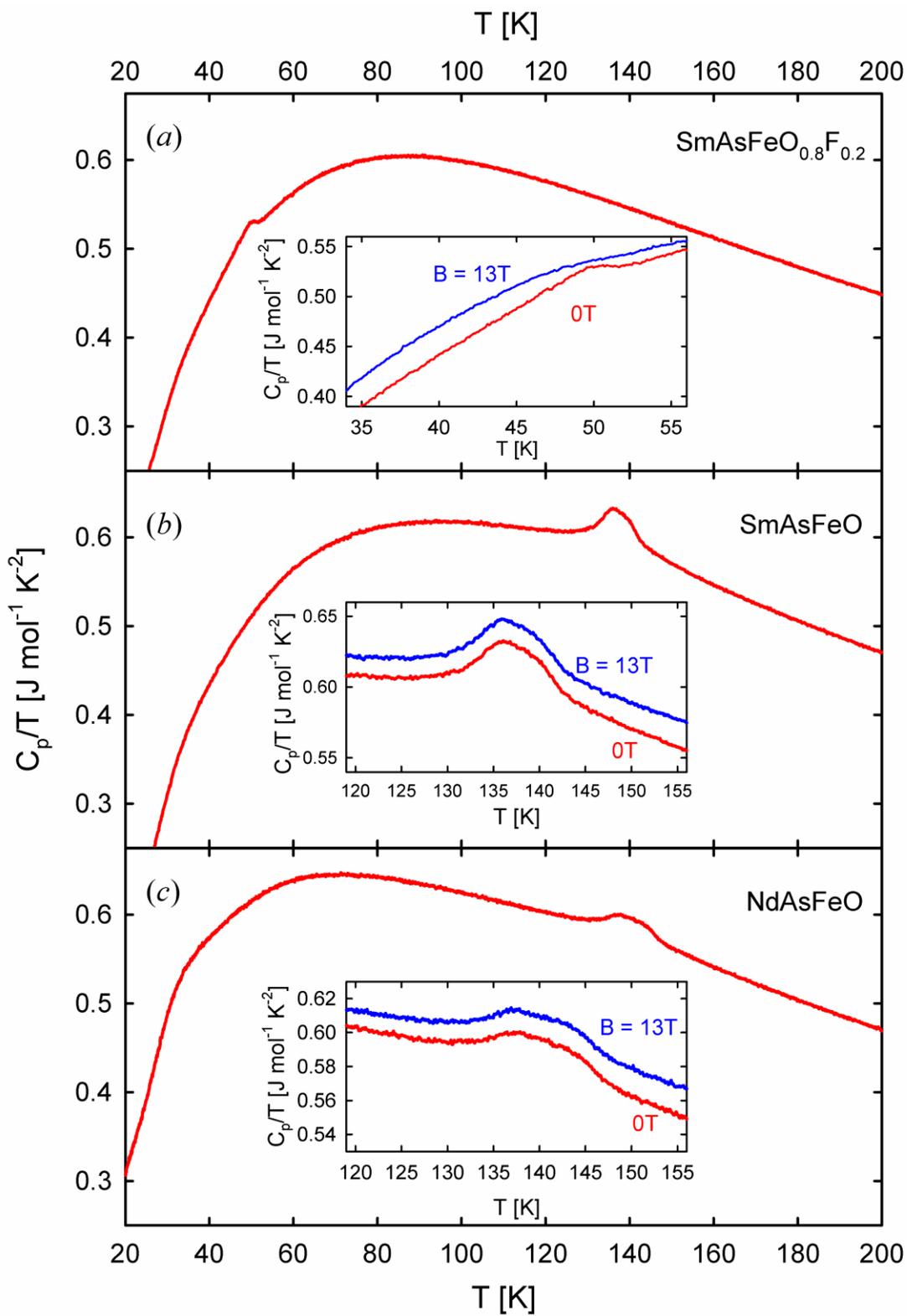

**Fig. 2**

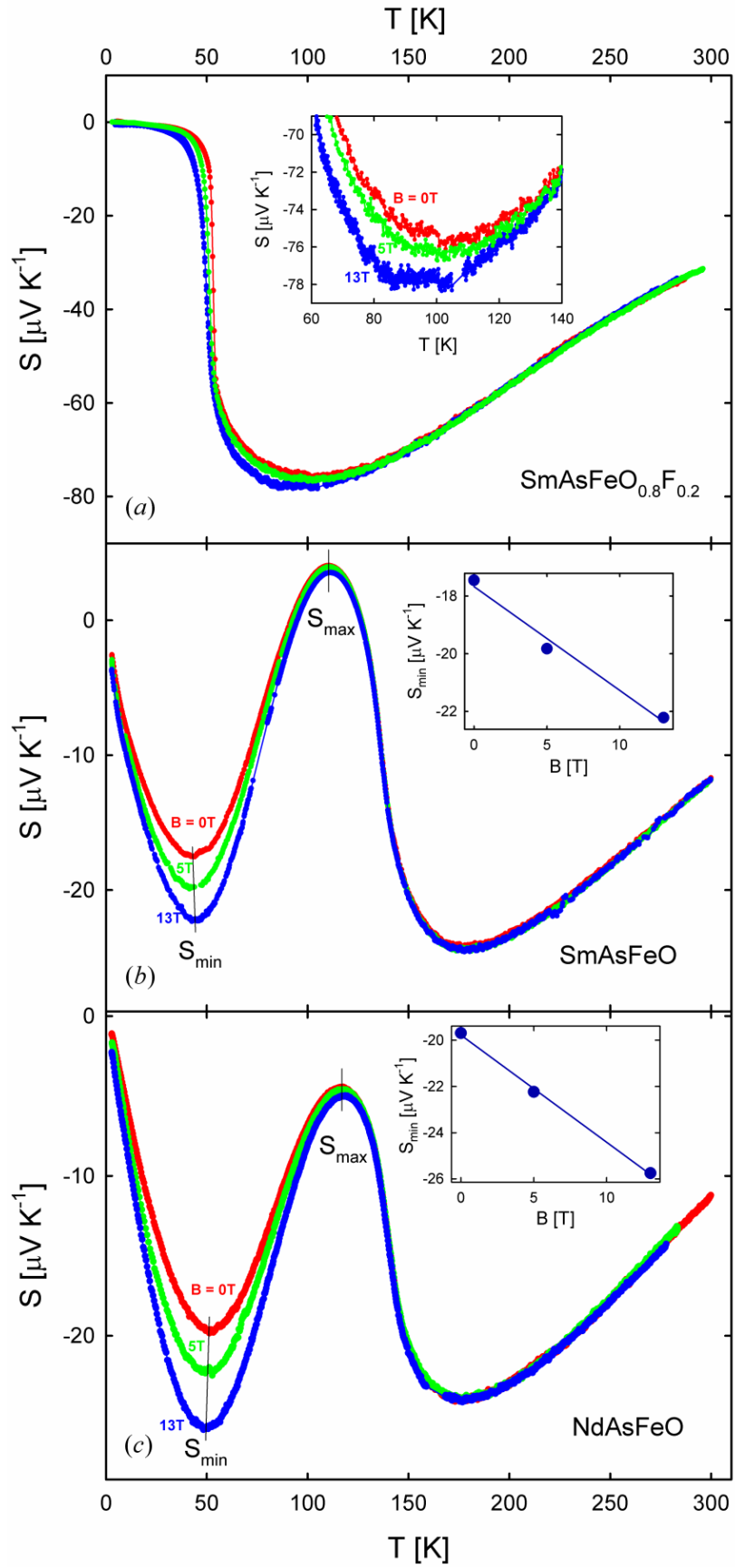

**Fig. 3**

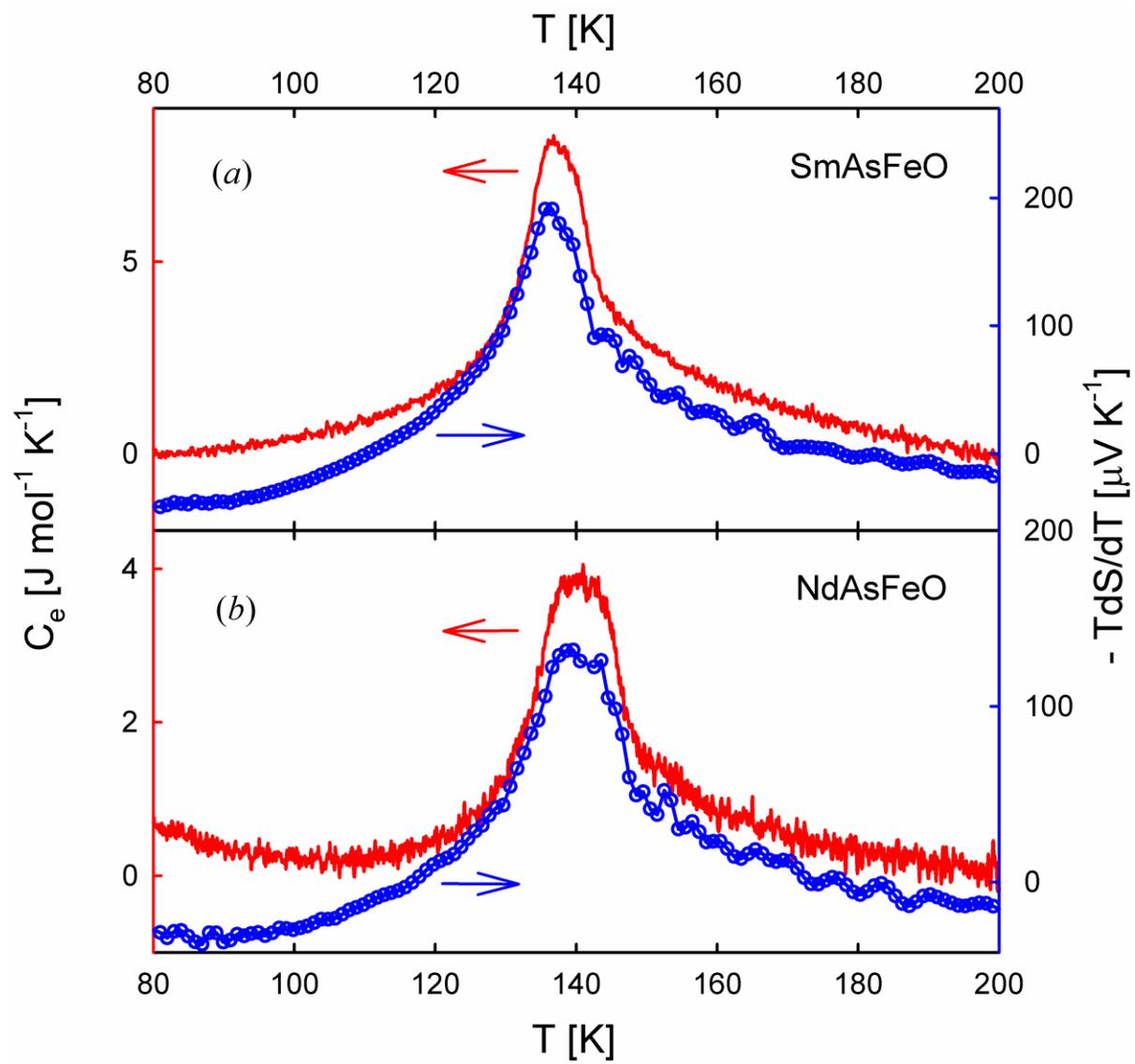

**Fig. 4**

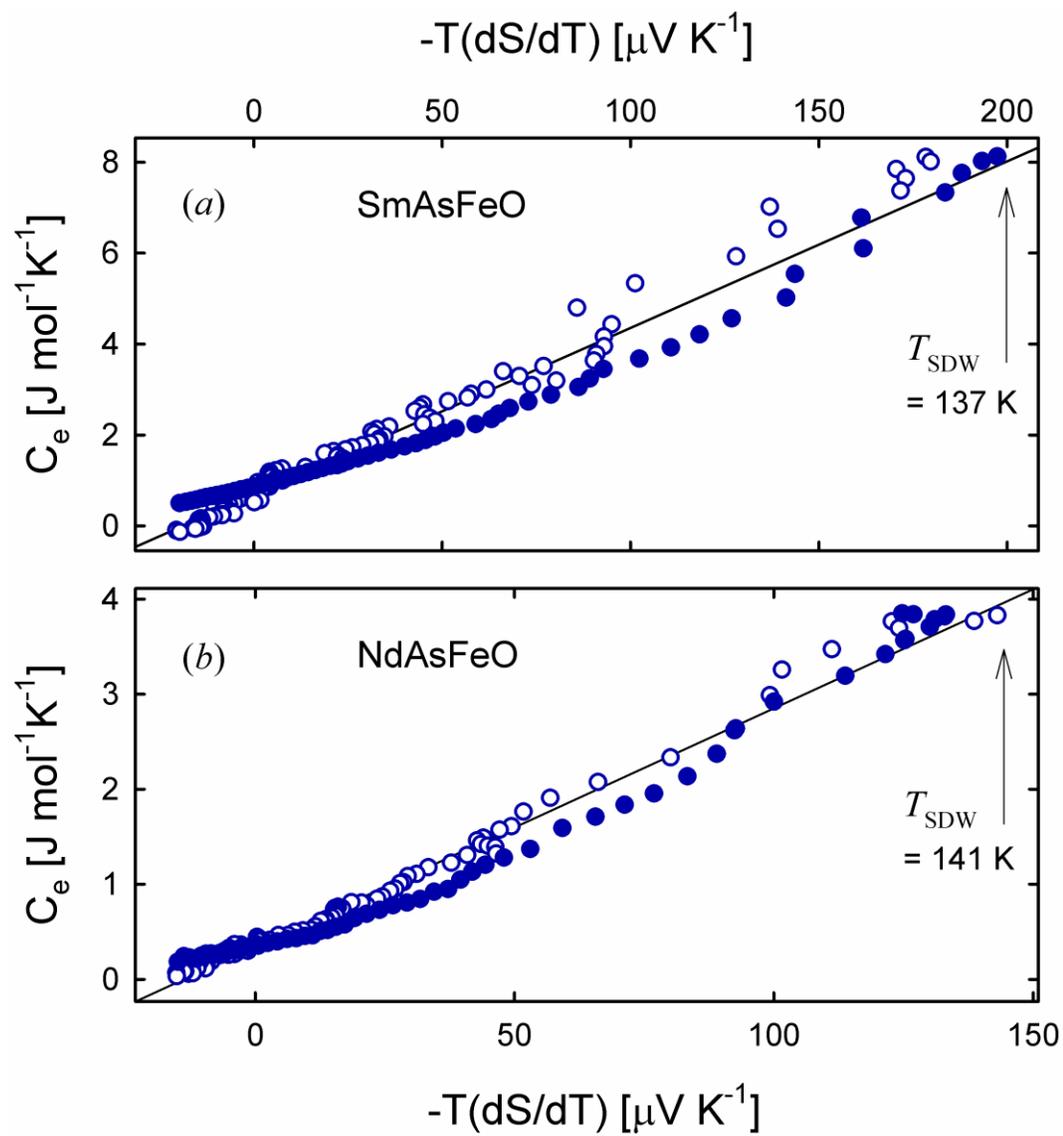

**Fig. 5**

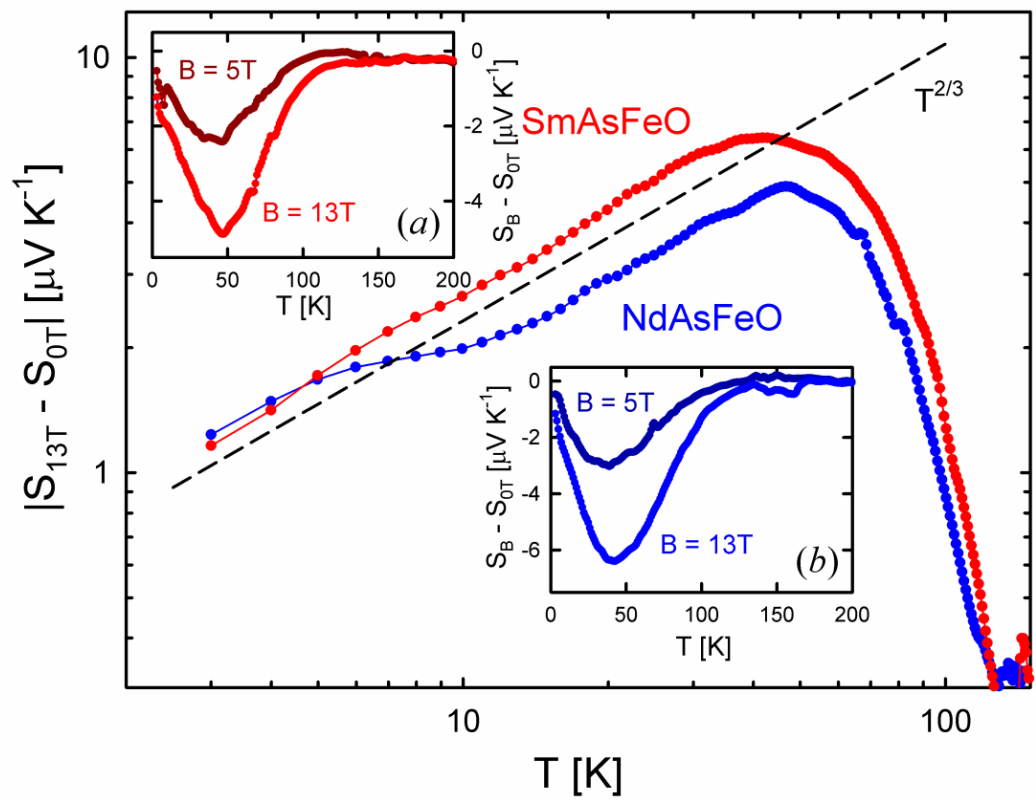